\documentclass{article}
\usepackage{spconf,amsmath,graphicx}

\usepackage{cite}
\usepackage{graphicx}
\usepackage{amssymb}
\usepackage{amsmath}
\usepackage{mathtools}
\usepackage{algorithm}
\usepackage{algpseudocode}
\usepackage{multirow}
\usepackage{adjustbox}
\usepackage{subfig}
\usepackage{makecell}
\usepackage{pbox}
\usepackage{bigstrut}
\usepackage{array}
\usepackage{cuted}
\usepackage{multirow} 
\usepackage{array}
\usepackage{amsthm}

\makeatletter
\renewcommand*\env@matrix[1][*\c@MaxMatrixCols c]{%
  \hskip -\arraycolsep
  \let\@ifnextchar\new@ifnextchar
  \array{#1}}
\makeatother

\title{Computationally iterative methods for salt-and-pepper denoising}
\name{ Jianwei Ke}
\address{University of Wisconsin-Madison\\
Department of Electrical and Computer Engineering\\ 
Madison, WI 53706, USA
}
\begin{document}
%
\maketitle
\begin{abstract}
Image restoration refers to the process of reconstructing noisy, destroyed, or missing parts of an image, which is an ill-posed inverse problem. A specific regularization term and image degradation are typically assumed to achieve well-posedness. Based on the underlying assumption, an image restoration problem can be modeled as a linear or non-linear optimization problem with or without regularization, which can be solved by iterative methods. In this work, we propose two different iterative methods by linearizing a system of non-linear equations and coupling them with the framework proposed in \cite{Chan2005_Newton}. The qualitative and quantitative experimental results demonstrate the correctness and efficiency of the proposed methods.
\end{abstract}
\section{Introduction}
\label{sec:intro}

Salt-and-pepper noise is a common type of impulse noise caused by malfunctioning pixels in image sensors \cite{2007caiCG}. For images contaminated by salt-and-pepper noise, the noisy pixels can either take the minimum value, i.e., 0, or the maximum value, i.e., 255, in a greyscale image. 

There are three popular non-learning types of approaches for removing salt-and-pepper noise. One type is based on median filter\cite{chen2001,Hwang2007, DSP}. This type of method can detect the noisy pixels accurately but do poorly in restoration when the noise ratio is high. These methods may lose details and cause distortion because the median of a selected window replaces the corrupted pixels. The second approach is a variational approach\cite{Nikolova2004}, which can preserve the details and the edges well, but the values of every pixel, including the uncorrected ones, are changed.  

The last approach is based on a two-phase scheme\cite{2007Cai,2007caiCG,Chan2005_Newton,Chan2005}. It combines the advantages of the previous two methods. In this first phase, an adaptive median method is used to detect the noisy pixels. In the second phase, the noisy pixels identified in the first phase are recovered by minimizing a special regularization functional, which boils down to solving a system of nonlinear equations. The details and edges are persevered in this method because of the special regularization functional, and the uncorrected pixels remain unchanged.  

For images corrupted by other noise like Gaussian noise, least-squares methods\cite{Vogel1998,Charbonnier1997} that utilize edge-preserving regularization functional can preserve the edges and details in images. However, these methods do not work in the presence of salt-and-pepper noise because they assume the Gaussianity of the underlying noise. Moreover, these methods will modify both the corrupted and non-corrupted pixels.

In this work, we proposed two novel iterative methods for restoring images based on the two-phase ideas in the presence of salt-and-pepper noise.


In section 2, we describe the problem formulation and the derivation for the key system of nonlinear equations. We present the two proposed approaches in section 3. In section 4, the experiment and results are shown. Finally, in section 5, we discuss and conclude the work. 

\section{Problem Formulation}
\label{sec:ProblemFormulation}
In the two-phase methods\cite{2007Cai,2007caiCG,Chan2005_Newton,Chan2005}, corrupted pixels are firstly identified by an adaptive median filter. Then in the second phase, the image is restored by minimizing a specialized regularization objective that applies only to those selected noise pixel candidates. Different schemes are applied to minimize the regularized functional in the second phase. For example, Chan \textit{et al.}\cite{Chan2005,Chan2005_Newton} proposed to use the variational methods and Newton's method. Cai et al.\cite{2007Cai, 2007caiCG} proposed using the Conjugate Gradient(CG) method and its variants for the minimization problem in the second phase of the algorithm.

\subsection{Noise detection by an adaptive median filter}
Let $\textbf{X}$ be a gray scale image of size $M \times N$. For $(i,j) \in \{1,\ldots,M\} \times \{1,\ldots,N\}$, $x_{i,j}$ denote the pixel at location $(i,j)$. Let $\mathcal{\phi}_{i,j}=\{x_{i,j-1}, x_{i,j+1},x_{i-1,j},x_{i+1,j}\}$ be the neighborhood of $x_{i,j}$. $s_{min}$, and $s_{max}$ denote the minimal and maximal value of all noisy pixels. In the classical salt-and-pepper noise, the observed noisy image, denoted by $\overline{\textbf{X}}$, is given by
\begin{align*}
    \overline{x}_{i,j} = \begin{cases}
        &s_{min}, \text{ with probability }p,\\
        &s_{max}, \text{ with probability }q,\\
        &x_{i,j}, \;\;\text{ with probability }1-p-q
    \end{cases}
\end{align*}
where $r = p+q$ denotes the noise level and $\overline{x}_{i,j}$ is the pixel of $\overline{\textbf{X}}$ at location $(i,j)$. The adaptive median filter guarantees that most of the salt-and-pepper noise is detected, provided that the window size is large enough. Notice that the noisy pixel candidates are replaced by the median $\overline{x}_{i,j}^{med,\omega}$ while the remaining pixels are left changed.

\subsection{Restoration of corrupted pixels}
Since all pixels in $\mathcal{N}$ are detected as noise contaminated, we only need to restore the pixels in $\mathcal{N}$ and keep those in $\mathcal{N}^c$ unchanged. For a noise candidate $x_{i,j}$, its neighborhood $\phi_{i,j}$ can be split two to groups $\phi_{i,j} =(\phi_{i,j} \cap \mathcal{N}^c) \cup (\phi_{i,j} \cap \mathcal{N})$. 
In \cite{Nikolova2004}, noisy pixel restoration is cast as a minimization of a functional of the following form, restricted to the noise candidate set $\mathcal{N}$:
\begin{equation}
    \label{eqn:init_cost}
    F_{\overline{\textbf{X}}, \alpha}(\textbf{u}) = \sum_{(i,j) \in \mathcal{N}} \left[ |u_{i,j} - \overline{x}_{i,j}| + \frac{\beta}{2}(S_{i,j}^1+S_{i,j}^2) \right] 
\end{equation}
where $\textbf{u} \in [0,255]^{|\mathcal{N}|}$, $|\mathcal{N}|$ is the cardinality of $|\mathcal{N}|$, 
\[S_{i,j}^1 = \sum_{(m,n) \in  \phi_{i,j} \cap \mathcal{N}^c}  2 \cdot \Phi_{\alpha}(u_{i,j} - \overline{x}_{m,n})\]  \[S_{i,j}^2 = \sum_{(m,n) \in  \phi_{i,j} \cap \mathcal{N}} \Phi_{\alpha}(u_{i,j} - u_{m,n})\]
$\Phi_{\alpha}(x)$ is chosen to be
\begin{align*}
    &\Phi_{\alpha}(x) = \sqrt{\alpha + x^2}, \alpha > 0
\end{align*}
It has been shown in \cite{Nikolova2004} that the data term $|u_{i,j} - \overline{x}_{i,j}|$ in $\eqref{eqn:init_cost}$ discourages those wrongly detected pixels in $\mathcal{N}$ from modified, but it also introduces a small bias on the restoration of corrupted pixels. However, in these two phase approach, all noisy pixels in detected at the first phase, so this term is unnecessary in the second phase. Hence, in \cite{Chan2005, Chan2005_Newton}, the first term is dropped and the minimization become
\begin{equation}
\label{eqn:cost}
    \mathcal{F}_{\overline{\textbf{X}}, \alpha}(\textbf{u}) = \sum_{(i,j) \in \mathcal{N}}(S_{i,j}^1 + S_{i,j}^2) 
\end{equation}
where $S_{i,j}^1, S_{i,j}^2$ are defined as above.

Now the problem remaining is to find an effective way to minimize the functional \eqref{eqn:cost}. Since $\mathcal{F}_{\overline{\textbf{X}}, \alpha}(\textbf{u})$ in \eqref{eqn:cost} is convex,  minimizing it is equivalent to to solving the following equation:
\begin{equation}
    \label{eqn:diff}
    \bigtriangledown  \mathcal{F}_{\overline{\textbf{X}}, \alpha}(\textbf{u}) = \left(S_{i,j}^{1'} + S_{i,j}^{2'} \right)_{(i,j) \in \mathcal{N}} = \textbf{0} 
\end{equation}
where $S_{i,j}^{1'}$ and $S_{i,j}^{2'}$ are the derivatives of $S_{i,j}^{1}$ and $S_{i,j}^{1}$ with respect to $u_{i,j}$. As in \cite{2007Cai}, we choose $\Phi_{\alpha}(x) = \sqrt{x^2 + \alpha}$. Hence, the system of non-linear equation for $\eqref{eqn:diff}$ becomes
\begin{align}
    \label{eqn:diff_spec}
    \bigtriangledown  \mathcal{F}_{\overline{\textbf{X}}, \alpha}(\textbf{u}) &= \left[\sum_{(m,n) \in  \phi_{i,j} \cap \mathcal{N}^c}  2 \cdot \Phi_{\alpha}'(u_{i,j} - \overline{x}_{m,n})\right]_{(i,j) \in \mathcal{N}}\\ \nonumber 
    & + \left[\sum_{(m,n) \in  \phi_{i,j} \cap \mathcal{N}} 2\cdot \Phi_{\alpha}'(u_{i,j} - u_{m,n})\right]_{(i,j) \in \mathcal{N}} = 0
\end{align}
where
\begin{equation}
    \label{eqn:diff2}
     \Phi_{\alpha}'(x) = x(\alpha + x^2)^{-\frac{1}{2}}
\end{equation}
It is worth noting that the Jacobian matrix of $\bigtriangledown  \mathcal{F}_{\overline{\textbf{X}}, \alpha}(\textbf{u})$ is symmetric.

In \cite{Chan2005, Chan2005_Newton}, a variational approach and Newton's Method with continuation are proposed to solve \eqref{eqn:diff_spec}. Conjugate Gradient(CG) based methods \cite{2007caiCG, 2007Cai, Zhang2014} are proposed to solve \eqref{eqn:diff_spec}. In this work, we propose two iterative methods coupled with the idea of Newton's method with continuation proposed in \cite{Chan2005_Newton}.

\section{Proposed methods}
The Newton's method update for \eqref{eqn:diff_spec} can be compactly written as 
\begin{equation}
    \textbf{u}_{n+1} = \textbf{u}_n - \textbf{J}_{ \bigtriangledown  \mathcal{F}}(\textbf{u}_n)^{-1} \bigtriangledown \mathcal{F}(\textbf{u}_n)
\end{equation}
where $\textbf{J}_{\bigtriangledown  \mathcal{F}}(\textbf{u}_n)$ is the Jacobian matrix of $\bigtriangledown  \mathcal{F}(\textbf{u}_n)$.

Rather than actually computing the inverse of the Jacobian matrix, it is usually preferred to solve the system of linear equations for efficiency and stability:
\begin{equation}
    \label{eqn:newtonSys}
    \textbf{J}_{\bigtriangledown  \mathcal{F}}(\textbf{u}_n)(\textbf{u}_{n+1} - \textbf{u}_n) = - \bigtriangledown  \mathcal{F}(\textbf{u}_n)
\end{equation}
Specifically, if we let $\textbf{z}_n = (\textbf{u}_{n+1} - \textbf{u}_n)$, then $\eqref{eqn:newtonSys}$ becomes
\begin{equation}
    \label{eqn:newtonSys2}
    \textbf{J}_{\bigtriangledown  \mathcal{F}}(\textbf{u}_n)\textbf{z}_n  = - \bigtriangledown  \mathcal{F}(\textbf{u}_n)
\end{equation}
\begin{equation}
    \textbf{u}_{n+1} = \textbf{u}_n + \textbf{z}_n
\end{equation}
Hence, Newton's method essentially boils down to solving a system of a linear equation. Here we propose two different ways to solve $\eqref{eqn:newtonSys}$. 

\subsection{Relaxation Coordinate Descend}
A common way to compute the optimal solution in a high dimensional space is coordinate descent optimization method, where one variable is optimized at a time while others remain fixed. So in Newton's update formula \eqref{eqn:newtonSys}, instead of computing the Jacobian matrix $\textbf{J}_{ \bigtriangledown  \mathcal{F}}(\textbf{u}_n)$ and the gradient vector $\bigtriangledown \mathcal{F}(\textbf{u}_n)$, we just need to compute the first $\bigtriangledown \mathcal{F}(u_{n,(i,j)})$ and the second derivative $\bigtriangledown^2 \mathcal{F}(u_{n,(i,j)})$ with respect to the variable currently being processed and update the variable in Newton's method with $\bigtriangledown \mathcal{F}(u_{n,(i,j)})$ and $\bigtriangledown^2 \mathcal{F}(u_{n,(i,j)}$. Namely, we relax the matrix version of Newton's method \eqref{eqn:newtonSys} to its scalar version:
\begin{equation}
    u_{n+1,(i,j)} = u_{n,(i,j)} - \frac{\bigtriangledown \mathcal{F}(u_{n,(i,j)})}{\bigtriangledown^2 \mathcal{F}(u_{n,(i,j)})}
\end{equation}
where $u_{n,(i,j)}$ is the element in $(i,j)$ position of $\textbf{u}_n$, 
\begin{align}
     \label{eqn:var_1diff}
     \bigtriangledown \mathcal{F}(u_{n,(i,j)}) 
     &= \sum_{(m,n) \in  \phi_{i,j} \cap \mathcal{N}^c} 2 \cdot \Phi_{\alpha}'(u_{i,j} - \overline{x}_{m,n})\\ \nonumber
     &+ \sum_{(m,n) \in  \phi_{i,j} \cap \mathcal{N}} 2\cdot \Phi_{\alpha}'(u_{i,j} - u_{m,n})
\end{align}
\begin{align}
     \label{eqn:var_2diff}
     \bigtriangledown^2 \mathcal{F}(u_{n,(i,j)}) &= \sum_{(m,n) \in  \phi_{i,j} \cap \mathcal{N}^c}  2 \cdot \Phi_{\alpha}''(u_{i,j} - \overline{x}_{m,n})\\ \nonumber
     &+ \sum_{(m,n) \in  \phi_{i,j} \cap \mathcal{N}} 2\cdot \Phi_{\alpha}''(u_{i,j} - u_{m,n})
\end{align}
The relaxation coordinate descend method is summarized in Algorithm \ref{alg:RCD}
\begin{algorithm}[H]
\begin{algorithmic}[0]
\caption{Relaxation Coordinate Descend}
\label{alg:RCD}
\State Initialize $\textbf{u}_0$ with the output of AMF
\For{$k=1,2,3,\ldots$}
\For{$u$ in $\textbf{u}_{k-1}$}
    \State Compute $\bigtriangledown \mathcal{F}(u_{k-1,(i,j)})$ and $\bigtriangledown^2 \mathcal{F}(u_{k-1,(i,j)})$ with $\eqref{eqn:var_1diff}$ and \eqref{eqn:var_2diff} respectively.
    \State $u_{k,(i,j)} = u_{k-1,(i,j)} - \frac{\bigtriangledown \mathcal{F}(u_{k-1,(i,j)})}{\bigtriangledown^2 \mathcal{F}(u_{k-1,(i,j)})}$
\EndFor
\EndFor
\end{algorithmic}
\end{algorithm}

\subsection{Minimum Residual Method(MINRES)}
Since the Jacobian matrix $ \textbf{J}_{\bigtriangledown  \mathcal{F}}(\textbf{u}_n)$ is symmetric, we can use the Minimum Residual algorithm(MINRES)\cite{MINRES} that requires the coefficient matrix to be symmetric to solve $\eqref{eqn:newtonSys}$. 
In CG methods, a breakdown can happen in the spirit of a zero pivot if the matrix is indefinite. Moreover, the minimization property of the CG methods is not well defined for indefinite matrices \cite{black_shirley}. 
The MINRES is a variant of the Lanczos method\cite{Lanczos1950} that supports the CG methods. Different from preconditioned conjugate gradient methods, MINRES allows the coefficient matrix to be indefinite. It voids the implicit LU factorization normally present in the Lanczos method, which is prone to breakdowns when zero pivots are encountered.

For the ease of discussion, we note \eqref{eqn:newtonSys2} as
\[\textbf{Ax = b}\]

In the conjugate gradient method, the residuals form a orthogonal basis for the space
\[ \textbf{\text{span}} = \{\textbf{r}^{(0)}, \textbf{Ar}^{(0)}, \textbf{A}^2\textbf{r}^{(0)} \} \]
If $\textbf{A}$ is not positive definite but symmetric, we can still construct an orthogonal basis for the Krylov subspace. Eliminating the search direction in CG ends up gicing a recurrence 
\begin{equation}
    \textbf{A}\textbf{r}^{(i)} = \textbf{r}^{(i+1)}\textbf{t}_{i+1,i} + \textbf{r}^{(i)}\textbf{t}_{i,j} + \textbf{r}^{(i-1)}\textbf{t}_{i-1,i}
\end{equation}
which can be written compactly in matrix form as 
\begin{equation}
    \textbf{A}\textbf{R}_i = \textbf{R}_{i+1} \overline{\textbf{T}}_i
\end{equation}
where $\overline{\textbf{T}}_i$ is a $(i+1)\times i$ tridiagonal matrix. Then one can find
\begin{equation}
    \textbf{x}^{(i)} \in \{ \textbf{r}^{(0)}, \textbf{A}\textbf{r}^{(0)}, \ldots, \textbf{A}^{i-1}\textbf{r}^{(0)} \}, \textbf{u}^{(i)} = \textbf{R}_i \overline{\textbf{y}}
\end{equation}
that minimizes 
\begin{equation}
    ||\textbf{A}\textbf{x}^{(i)} - \textbf{b}||_2 = ||\textbf{A}\textbf{R}_i \overline{\textbf{y}} - \textbf{b}||_2 = || \textbf{R}_{i+1}\overline{\textbf{T}}\textbf{y} - \textbf{b} ||_2 
\end{equation}
Then we obtain the final expression:
    \begin{equation}
        \label{eqn:minres}
        ||\textbf{A}\textbf{x}^{(i)} - \textbf{b}||_2 = || \textbf{D}_{i+1} \overline{\textbf{T}}_i \textbf{y} - ||\textbf{r}||^{(0)}_2 \cdot \textbf{e}^{(1)}  ||_2
    \end{equation}
Then the final expression can be viewed as a minimum norm least squares problem by utilizing the fact that if $\textbf{D}_{i+1} = diag(||\textbf{r}^{(0)}||_2, ||\textbf{r}^{(1)}||_2,\ldots,||\textbf{r}^{(i)}||_2 )$, then $\textbf{R}_{i+1}\textbf{D}^{-1}_{i+1}$ is an orthonormal transformation with respect to the current Krylov subspace. The element $(i+1,i)$ position of $\overline{\textbf{T}}_i$ can be destroyed by a simple rotation, and the resulting upper bidiagonal system can be solved efficiently, leading to the MINRES method.

The MINERS method is summarized in Algorithm
\begin{algorithm}[H]
\begin{algorithmic}[0]
\caption{Minimum Residual Method(MINRES)}
\label{alg:MINERS}
\State Initialize $\textbf{u}_0$ with the output of AMF
\State $r_0 = f - Au_0$
\State $q_1 = \frac{r_0}{||r_0||_2}$
\For{$k=1,2,3,\ldots$}
\State $v = Aq_k$
\For{$i = 1:k$}
    \State $h_{ik} = q_{i}^Tv$
    \State $v =v -h_{ik}q_i$ \Comment{orthonogilize to previous vectors}
\EndFor
\State $h_{k+1,k} = ||v_2||$, $q_{k+1} = \frac{v}{h_{k+1,k}}$
\State Solve \eqref{eqn:minres}
\EndFor
\end{algorithmic}
\end{algorithm}

\subsection{Newton's method with continuation}
It is pointed out in \cite{Chan2005_Newton} that the potential function $\Phi_{\alpha}(x)$ should be close to $|x|$ while maintaining to be differentiable at zero to avoid stair-caising effect. 

Hence in $\Phi_{\alpha}(x) = \sqrt{\alpha + x^2}$, $\alpha$ should be close to 0, but if $\alpha$ is set this way, $\Phi_{\alpha}'(x)$ will have a sharp increase near the solution, and Newton's method will diverge easily. To remedy this situation, an iterative framework coupled with Newton's method is proposed in \cite{Chan2005_Newton}, where in each iteration, a Newton' method is applied to solve \eqref{eqn:diff_spec}, and the solution is used as the initial guess for the next iteration with a larger $\alpha$. We couple this idea with our two proposed methods. 


\section{Experiments and Results}
\subsection{Setup}
We use Python 3.0 on a desktop with an Intel Core i9 11900k 4.8 GHz CPU with a 32 G memory and 16 MB Cache throughout the experiment. The test images include \textit{lena, cameraman, barbara, boat, pirate, woman-darkhair}, all of which have 512x512 resolution. We are mainly concerned restoration quality and processing time of each method. To access the restoration performance qualitatively, we use the PSNR(peak signal to noise ratio) defined as 
\begin{equation}
    PSNR(F_{k+1},F_{k}) = 10 \log_{10} \left( \frac{peakVal^2}{MSE(I^r,I)} \right)
\end{equation}
where $peakval$ is the maximum pixel value and $MSE(F_{k+1},F_{k})$ the mean square error between the reference (original) image $I^r$ and the reconstructed image $I$. Since we are concerned with the processing time of solving the minimization problem in the second phase, only the time for the second phase is reported. Each experiment was repeated five times, and we reported the average.

The stopping criteria of all methods is set to 
\[ \frac{|| \textbf{u}^k - \textbf{u}^{k-1} ||}{||\textbf{u}^k||} \leq 10^{-4} \text{ and } \frac{|\mathcal{F}_{\alpha}(\textbf{u}^k) - \mathcal{F}_{\alpha}(\textbf{u}^{k-1}) |}{\mathcal{F}_\alpha(\textbf{u}^k)} \leq 10^{-4} \]
or the number of iteration has reached $ite_{max} = 500$.

\begin{table}[!t]
\renewcommand{\arraystretch}{1.3}
\centering
\caption{Averaged Timing of various algorithms for all testing images in minutes}
\label{tab:time}
\begin{tabular}
{|p{1cm} |c |c | c | c | c |c}\hline Noise Ratio & FR\cite{FR} & PR\cite{PR} & HS\cite{HS} & Relax & Minres\\
\hline 30\% & 1.07 & 0.89 & 0.86 & \textbf{0.20} & 0.91\\
\hline 50\% & 2.29 & 1.79 & 1.86 & \textbf{0.34} & 4.33\\
\hline 70\% & 4.95 & 3.59 & 3.82 & \textbf{0.48} & 13.09\\
\hline 90\% & 11.72 & 8.23 & 8.40 & \textbf{0.63} & 28.24\\
\hline
\end{tabular}
\end{table}

\begin{table}[!t]
\renewcommand{\arraystretch}{1.3}
\centering
\caption{Averaged PSNR of various algorithms for all testing images image in dB}
\label{tab:PSNR}
\begin{tabular}
{|p{1cm}| c | c | c | c | c |c}\hline Noise Ratio & FR\cite{FR} & PR\cite{PR} & HS\cite{HS} & Relax & Minres\\
\hline 30\% & 38.20 & 38.16 & \textbf{38.21} & 38.20 & 38.17\\
\hline 50\% & 34.45 & 34.42 & 34.44 & \textbf{34.66} & 34.33\\
\hline 70\% & 31.11 & \textbf{31.18} & 31.14 & 31.01 & 31.06\\
\hline 90\% & 11.72 & 26.26 & \textbf{26.47} & 25.89 & 26.33\\
\hline
\end{tabular}
\end{table}

\subsection{Comparison of the CG Type Methods and the proposed methods}
We compare the three nonlinear CG type methods in \cite{FR, PR, HS}, denoted as FR, PR, and HS respectively.

In Table \ref{tab:time} and Table \ref{tab:PSNR}, we report the averaged running time (in minutes) and PNSR (in dB) for the three non-linear CG methods with various noise ratios and the two proposed methods. We see that PR is the most efficient among the three non-linear CG methods.

For Newton's method with continuation, we choose $\alpha$ to be 
\begin{align*}
    &160000 \xrightarrow{\div \text{32}} 5000 \xrightarrow{\div \text{4}} 1250 \xrightarrow{\div \text{4}} 312.5 \xrightarrow{\div \text{2}} 156.25 \xrightarrow{\div \text{2}} 78.125\\ 
    &\xrightarrow{\div \text{2}} 39.0625 \xrightarrow{\div \text{2}} \cdots
\end{align*}

Similar to the nonlinear CG methods, we set the maximum iterations for the two proposed methods to 500 and the error tolerance is $10^{-4} $. As we can see the relaxation method just takes less than 1 min in all cases. The main reason is that it avoids evaluating the Jacobian matrix $\textbf{J}_{\bigtriangledown \mathcal{F}}$ and only requires evaluating scalar functions. 

Overall the three non-linear CG methods have similar performance and outperform the Minres-based Newton's method with continuation in efficiency and restoration quality. However, the proposed relaxation method with continuation has similar performance as the non-linear CG methods but is more efficient than all the CG type methods.
\begin{figure}%
    \centering
    \subfloat{{\includegraphics[ scale=0.5,trim=4cm 8cm 3cm 8.2cm, clip]{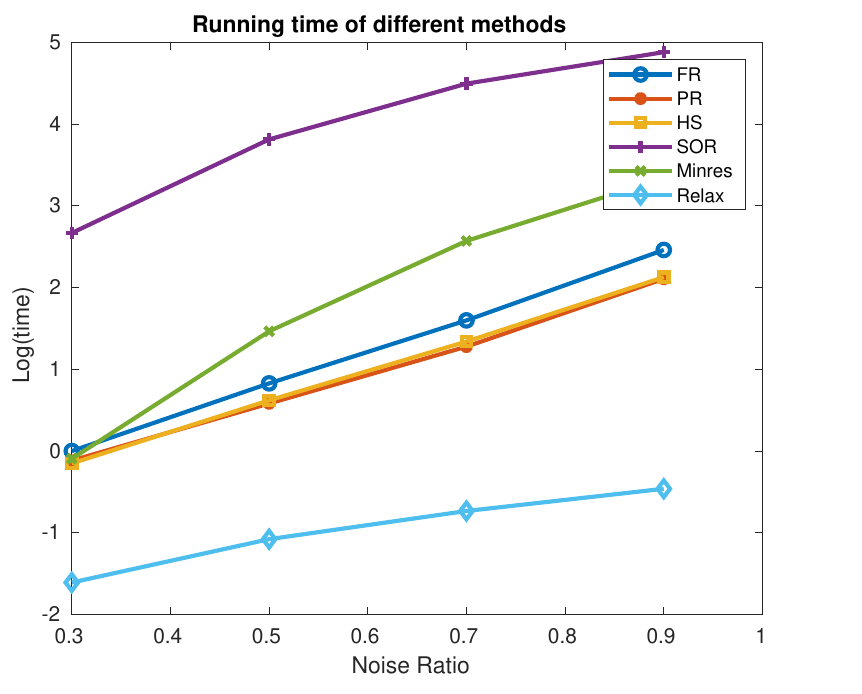} }}%
    \quad
    \subfloat{{\includegraphics[ scale=0.5,trim=4cm 8cm 3cm 8.2cm, clip]{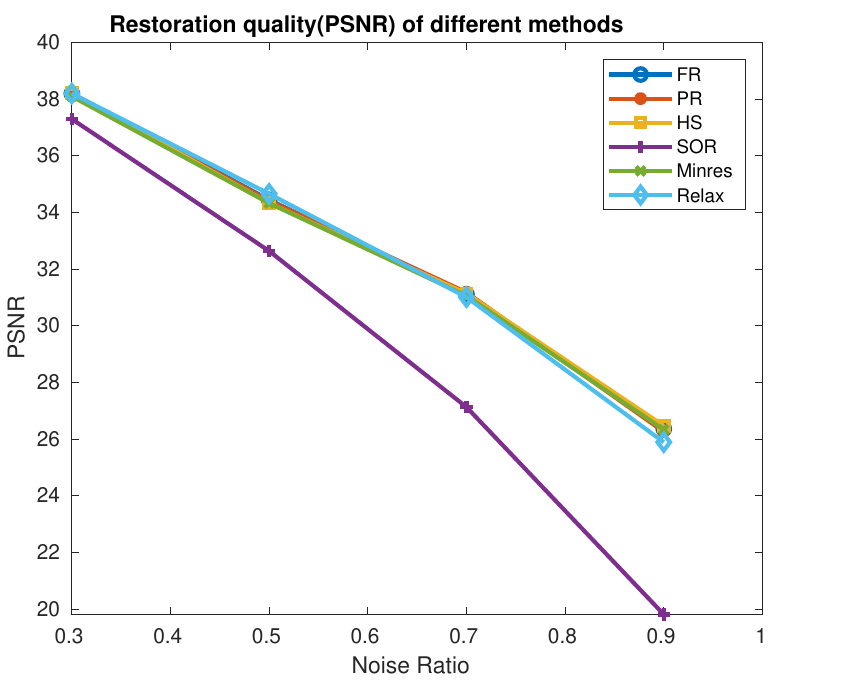} }}%
    
    \caption{Comparison of different methods in terms of running time and PSNR.}
    \vspace{-0.5cm}
     \label{fig:comp}
\end{figure}

\section{Discussion and Conclusion}
In this paper, we proposed two computationally iterative methods based on the two-phase methods \cite{2007caiCG,2007Cai,Chan2005_Newton} for restoring image corrupted by salt-and-pepper noise, wherein the first phase, the corrupted pixels are identified by an adaptive median filter. Then in the second phase, the image is restored by minimizing a specialized regularization objective that applies only to those selected noise pixel candidates. We linearize the key system of non-linear equations based on Newton's method and propose two ways to optimize the new objective within the framework of \cite{Chan2005_Newton}. Experiment results confirm the correctness and efficiency of the two proposed methods, which can be critical to downstream Computer Vision tasks such as image-based 3D reconstruction \cite{Ke2022} and video stabilization\cite{Ke2023}. 


The performance of each algorithm, especially the running time, heavily relies on the specific implementation. In current implementation, evaluating the cost function $\mathcal{F}$, it gradient $\bigtriangledown \mathcal{F}$, and the Jacobian matrix $\textbf{J}_{\bigtriangledown \mathcal{F}}$ are done separately. Since they have similar structures, a better implementation is to evaluate the three quantities jointly. 

\newpage
\bibliographystyle{IEEEbib}
\bibliography{main}

\end{document}